\documentclass{article}

\usepackage{arxiv}

\usepackage[utf8]{inputenc} 
\usepackage[T1]{fontenc}    
\usepackage{hyperref}       
\usepackage{url}            
\usepackage{booktabs}       
\usepackage{amsfonts}       
\usepackage{nicefrac}       
\usepackage{microtype}      
\usepackage{lipsum}		
\usepackage{subcaption} 
\usepackage{graphicx}
\usepackage{natbib}
\usepackage{doi}
\usepackage{amssymb}
\usepackage{amsmath}
\usepackage{amsthm}
\usepackage{mathtools}
\usepackage{bm}
\usepackage{dirtytalk}
\usepackage{verbatim}
\usepackage{parskip}
\usepackage{dsfont}
\usepackage{algorithm}
\usepackage{algpseudocode}
\usepackage[dvipsnames]{xcolor}
\usepackage[normalem]{ulem}

\usepackage{float}
\newcommand{\R}{\mathbb{R}} 
\newcommand{\C}{\mathbb{C}}

\newcommand{\F}{\mathcal{F}}

\newcommand{\bk}{\bm{k}}

\newcommand{\bchi}{\bm{\chi}}

\newcommand{\br}{\bm{r}}

\newcommand{\norm}[1]{\|#1\|}

\newtheorem{theorem}{Theorem}

\newtheorem{proposition}{Proposition}
\newtheorem{corollary}{Corollary}

\newcommand{\TE}{\mathrm{TE}}
\newcommand{\1}{\mathbf{1}}
\renewcommand{\Re}{\operatorname{Re}}
\renewcommand{\Im}{\operatorname{Im}}
\newcommand{\clip}{\operatorname{clip}}
\newcommand{\wrap}{\operatorname{wrap}}
\newcommand{\tr}{\operatorname{tr}}
\newcommand{\Hm}{\mathsf{H}}            
\newcommand{\bbeta}{\bm{\beta}}
\newcommand{\brho}{\bm{\rho}}
\newcommand{\bdelta}{\bm{\delta}}
\newcommand{\btheta}{\bm{\theta}}
\newcommand{\Tri}{\mathcal{T}}

\newcommand{\chiqsm}{\chi^{\mathrm{QSM}}}

\title{Susceptibility source separation via Closed-form Alternating Newton-type Optimization}
\author{ \href{https://orcid.org/0009-0005-4522-0343}{\includegraphics[scale=0.06]{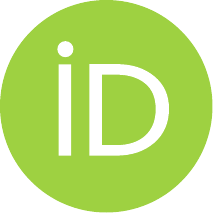}\hspace{1mm}Ignacio Contreras-Z\'u\~niga}\\
	Biomedical Imaging Center\\
	Pontificia Universidad Católica de Chile\\
	\texttt{iscontreras@uc.cl} \\
	\And
	Cristian Tejos \\
	Department of Electrical Engineering \\
    Biomedical Imaging Center\\
    Millennium Institute for Intelligent Healthcare Engineering\\
	Pontificia Universidad Católica de Chile \\
	\texttt{ctejos@uc.cl} \\
	\And
	 \href{https://orcid.org/0000-0002-1196-6703}{\includegraphics[scale=0.06]{orcid.pdf}\hspace{1mm}Carlos Milovic} \\
	Department of Electrical Engineering \\
    Biomedical Imaging Center\\
	Pontificia Universidad Católica de Chile \\
	\texttt{cmilovic@uc.cl} \\
}

\renewcommand{\shorttitle}{Susceptibility Source Separation via CANON-QSM}

\hypersetup{
pdftitle={Susceptibility source separation via CANON-QSM},
pdfsubject={physics.med-ph, eess.SP},
pdfauthor={Contreras-Z\'u\~niga, Cristian Tejos, Carlos Milovic},
pdfkeywords={Quantitative Susceptibility Mapping, Susceptibility Source Separation, Inverse Problems},
}

\begin{document}
\maketitle

\begin{abstract}
Quantitative Susceptibility Mapping (QSM) provides a bulk magnetic susceptibility estimate per voxel, so contributions from paramagnetic and diamagnetic sources within the same voxel partially cancel. Susceptibility source separation methods identify these compartments by fitting a three-pool complex exponential model to multi-echo gradient-echo (GRE) data. However, conventional voxel-wise fitting with general-purpose constrained solvers requires several hours per whole-brain volume and often fails in high-susceptibility regions. We present CANON-QSM (Closed-form Alternating Newton-type Optimization for QSM), a solver that reformulates the subproblems of the three-pool model and solves them alternately via exact closed-form updates and  Newton-type steps, yielding a substantial per-voxel computational speedup. Both in silico and in vivo evaluations demonstrate not only a massive reduction in reconstruction time, completing a whole-brain volume in under two minutes, but also improved accuracy and robustness against artifacts in high-susceptibility regions. Finally, an echo-time analysis of the three-pool model shows that source separation is driven by second-order nonlinearities in signal decay and phase evolution rather than first-order bulk estimates.
\end{abstract}


\section{Introduction}\label{sec:introduction}

Quantitative susceptibility mapping (QSM) estimates the magnetic susceptibility of tissue from the phase of gradient-echo (GRE) MRI~\cite{brown2014magnetic,qsm2024recommended,choi2014inverse} and is widely used as a biomarker of brain iron and myelin in aging and neurodegenerative diseases. Standard QSM calculates a single bulk susceptibility value per voxel under the assumption that the local phase evolves linearly with echo time ($\TE$). In brain parenchyma, however, paramagnetic iron and diamagnetic myelin frequently coexist within the same voxel. Because their induced frequency shifts have opposite signs, they partially cancel, so the net bulk susceptibility can remain close to zero even when both sources are abundant, and a change in bulk susceptibility cannot be uniquely attributed to iron or myelin without additional information.

Susceptibility source separation resolves this ambiguity by exploiting the fact that both source types dephase spins and accelerate transverse relaxation, whereas their frequency shifts have opposite signs. For example, $\chi$-separation combines $R_2^*$ and QSM with $R_2$ or $R_2'$ maps~\cite{shin2021chi}, requiring an auxiliary spin-echo acquisition that increases scan time. Alternatively, DECOMPOSE-QSM~\cite{chen2021decompose} relies solely on multi-echo GRE data, modeling the complex signal of each voxel as the sum of paramagnetic, diamagnetic, and neutral pools with decay rates and frequency shifts derived from static dephasing theory~\cite{yablonskiy1994theory}.

A major practical obstacle to GRE-only source separation is its computational cost. DECOMPOSE-QSM fits each voxel by alternating over three parameter blocks, solving each block iteratively with a general-purpose constrained MATLAB solver~\cite{chen2021decompose}. This procedure requires several hours per brain~\cite{otsuka2026deep}, and fitting artifacts have been reported in large cerebral veins~\cite{fuchs2024practical}. Although a deep-learning approximation has recently been proposed to accelerate inference~\cite{otsuka2026deep}, network predictions are not guaranteed to satisfy the physical forward model.
In this work, we solve the three-pool model directly by reformulating each subproblem. Specifically, the pool-concentration and baseline-relaxation blocks admit exact, iteration-free closed-form solutions, while the susceptibility block is formulated using a phase residual that is insensitive to $2\pi$ wrapping, paired with an analytic Jacobian and closed-form $2\times2$ Newton-type (Levenberg-Marquardt) steps with a parallel multi-start strategy. We term this method Closed-form Alternating Newton-type Optimization for QSM (CANON-QSM). Because all subproblems reduce to elementary per-voxel operations, the solver executes voxel-parallel on a GPU, providing a massive computational speedup. We validate CANON-QSM on a realistic numerical brain phantom and \textit{in vivo}, achieving whole-brain susceptibility separation in under two minutes.

Furthermore, we bridge a theoretical gap in the literature from a signal-processing perspective by expanding the three-pool complex signal in powers of echo time. We show that linear-phase QSM and mono-exponential $R_2^*$ fitting capture only first-order echo-time quantities, whereas paramagnetic and diamagnetic sources are separated by second-order effects: the echo-time drift of the apparent susceptibility and the curvature of the log-magnitude decay. This analysis explains why per-echo QSM reconstructions are required as input.

\textbf{Notation.} All quantities refer to a single voxel unless stated otherwise. Let $N_e$ denote the number of echoes, indexed by $e$, and let $\mathbf{t}=(\TE_1,\dots,\TE_{N_e})^\top$ with $0<\TE_1<\dots<\TE_{N_e}$. The symbols $\odot$ and $\oslash$ denote elementwise multiplication and division, and $\exp$, $\log$, $|\cdot|$, and $\arg$ act elementwise, with $\arg\in(-\pi,\pi]$. Here, $\mathbf{v}^\Hm$ denotes the conjugate transpose, $\bar{\mathbf{v}}$ the complex conjugate, $\1$ a vector of ones, and $\clip_{[l,u]}(x)=\min\{\max\{x,l\},u\}$.

\subsection{GRE signal and dipole deconvolution}
Let $\Omega\subset\R^3$ denote the imaging domain placed in a main static magnetic field $\bm{B}_0 = B_0\widehat{\bm{b}}_0$ with unit direction $\widehat{\bm{b}}_0=(0,0,1)^\top$ and gyromagnetic ratio $\gamma$. In standard QSM, a voxel at spatial position $\br\in\Omega$ is characterized by a bulk magnetic susceptibility distribution $\chi(\br)$ (in ppm) and an effective transverse relaxation rate $R_2^*(\br)\ge 0$. Under the linear mono-exponential model, the continuous-time complex GRE signal at echo time $t \ge 0$ is given by~\cite{salomir2003fast, marques2005application}
\begin{equation}\label{eq:gre_bulk}
S_{\mathrm{GRE}}(\br,t) = M_0(\br)\,e^{-R_2^*(\br)\,t}\,e^{\,i\bigl(\phi_0(\br) + \gamma\,B(\br)\,t\bigr)},
\end{equation}
where $M_0(\br) \ge 0$ is the equilibrium spin-density magnitude at $t=0$, $\phi_0(\br)\in(-\pi,\pi]$ is the transmit/receive radiofrequency phase offset, and $B(\br)$ is the $z$-component of the total magnetic field perturbation. Consequently, the unwrapped phase $\phi(\br,t)$ of $S_{\mathrm{GRE}}(\br,t)$ evolves linearly with time according to
\begin{equation}\label{eq:phase_map}
\phi(\br,t) = \phi_0(\br) + 2\pi\, f(\br)\,t, \qquad f(\br) = \frac{\gamma}{2\pi}\, B(\br),
\end{equation}
where $f(\br)$ (or, equivalently, the relative field shift $\psi_{\text{total}}(\br) = B(\br)/B_0$) is the total frequency shift. This perturbation splits into a harmonic background field $B_{\mathrm{bg}}(\br)$, generated by susceptibility sources outside the region of interest $\Omega$ (such as air-tissue interfaces) that satisfies $\Delta B_{\mathrm{bg}} = 0$ in $\Omega$, and a local tissue field $B_{\mathrm{loc}}(\br)$ induced by tissue magnetization within $\Omega$.

Under the Lorentzian sphere approximation, the relative local field perturbation $\psi(\br) :=  B_{\mathrm{loc}}(\br)/B_0$ is linked to the spatial susceptibility distribution $\chi(\br)$ via three-dimensional convolution with the unit magnetic dipole kernel $d(\br)$~\cite{brown2014magnetic,choi2014inverse}:
\begin{equation}\label{eq:dipole_fourier}
\F(\psi)(\bk) = \,D(\bk)\,\F(\chi)(\bk), \qquad D(\bk) = \frac{1}{3} - \frac{(\widehat{\bm{b}}_0\cdot\bk)^2}{\norm{\bk}^2}\quad (\bk\neq\bm{0}),
\end{equation}
with $D(\bm{0})=0$, where $\F$ denotes the continuous 3D Fourier transform. Standard QSM estimates a single scalar $\chi(\br)$ per voxel by unwrapping $\phi(\br,t)$, removing $\phi_0(\br)$ and $B_{\mathrm{bg}}(\br)$, and solving the ill-posed dipole deconvolution problem, whose Fourier kernel vanishes on the magic-angle cone.

The single-compartment model in~\eqref{eq:gre_bulk} assumes a homogeneous sub-voxel frequency shift and a mono-exponential magnitude decay. In brain tissue, however, a single voxel typically contains water protons surrounding two distinct populations of sub-voxel magnetic perturbers: paramagnetic inclusions, dominated by iron stored in ferritin, and diamagnetic inclusions, dominated by myelin. Each inclusion distorts the microscopic magnetic field within the voxel. Protons precessing near an inclusion experience local field offsets and lose phase coherence, accelerating the transverse decay of the voxel signal, while the net microscopic field offset shifts its mean frequency. Because paramagnetic and diamagnetic inclusions induce frequency shifts of opposite signs that partially cancel in the net phase~\eqref{eq:phase_map} while both accelerate $R_2^*$ relaxation, the composite voxel signal departs from the mono-exponential model in~\eqref{eq:gre_bulk}.

\section{Theory}\label{sec:theory}

\subsection{Sub-voxel Susceptibility and Signal Model}\label{sec:theorySignal}

Following~\cite{chen2021decompose}, at each voxel we model the signal as the sum of three pools: paramagnetic ($+$), diamagnetic ($-$), and neutral ($0$). We aim to recover the concentrations $\mathbf{c}=(c^+,c^-,c^0)^\top$, which represent the signal fractions at $t=0$ contributed by each pool, satisfying $c^k\ge0$ and $c^++c^-+c^0=1$, the intrinsic susceptibilities $\chi^+\ge0$ and $\chi^-\le0$ (in ppm), collected in $\bchi=(\chi^+,\chi^-)^\top$, and the baseline relaxation rate $R\equiv R^*_{2,0}$ of the neutral medium. In the static dephasing regime, where water diffusion during the echo train is negligible relative to the spatial scale of the field perturbations, a susceptibility source $\chi$ adds a transverse decay rate $a|\chi|$ and a frequency shift $p\chi$ to the signal of its pool~\cite{yablonskiy1994theory}.
 
The complex GRE signal of the voxel at time $t$ after excitation is therefore a sum of complex exponentials~\cite{chen2021decompose}:
\begin{equation}\label{eq:signalCanon}
S(t)=c^+e^{-(a\chi^++R)\,t+ip\chi^+t}+c^-e^{-(-a\chi^-+R)\,t+ip\chi^-t}+c^0e^{-Rt},
\end{equation}
where
\begin{equation}\label{eq:signalConstants}
a=\tfrac{2\pi}{9\sqrt3}\gamma B_0\cdot10^{-6},\qquad p=\tfrac23\gamma B_0\cdot10^{-6}.
\end{equation}
At 3\,T, $a\approx323.5$\,s$^{-1}$ppm$^{-1}$ and $p\approx535.0$\,rad\,s$^{-1}$ppm$^{-1}$. Since $\chi^+\ge0\ge\chi^-$, both $a\chi^+$ and $-a\chi^-$ equal $a|\chi^\pm|$, thus, the decay rate reports the magnitude of each perturbation and the frequency shift reports its sign. Both sources accelerate transverse decay, but they rotate the phase in opposite directions. The model has five independent real parameters, and each echo provides two real measurements, so at least three echoes are required.

We combine both effects of each source into the complex coefficients
\begin{equation}\label{eq:u}
u^+=-a+ip,\qquad u^-=a+ip,
\end{equation}
so that $u^\pm\chi^\pm=-a|\chi^\pm|+ip\chi^\pm$, and~\eqref{eq:signalCanon} becomes
\begin{equation}\label{eq:signalCanon2}
S(t)=e^{-Rt}\big(c^+e^{u^+\chi^+t}+c^-e^{u^-\chi^-t}+c^0\big).
\end{equation}
The factor $e^{-Rt}$ is shared by all pools, and the normalized model satisfies $S(0)=1$. 

The measured GRE phase also contains non-local phase contributions that~\eqref{eq:signalCanon2} does not model, such as background fields from air-tissue interfaces outside the voxel and receiver phase offsets—so fitting the model directly to the raw phase would falsely attribute these terms to $\chi^\pm$. As in~\cite{chen2021decompose}, the measured phase is therefore replaced by the phase induced by the locally reconstructed susceptibility, yielding the synthesized local signal
\begin{equation}\label{eq:signalSynth}
y(t)=\frac{M(t)}{M_0}\,e^{\,ip\,\chiqsm(t)\,t},
\end{equation}
where $M(t)$ is the measured GRE magnitude and $\chiqsm(t)$ is the QSM map reconstructed from the echo acquired at time $t$. This procedure removes background fields, inverts the dipole convolution~\eqref{eq:dipole_fourier}, and re-encodes the result via~\eqref{eq:signalSynth} using the same scaling constant $p$, preserving the echo-time phase dependence exhibited by mixed voxels. The normalization factor $M_0$ is extrapolated per voxel so that $|y(t)|\to1$ as $t\to0$, matching $S(0)=1$: $M_0=\exp(\widehat\mu+\widehat R^*_2\,\overline{\TE})$ is the $t=0$ intercept of a log-linear fit of $M(t)$, where $\widehat\mu$ is the mean log-magnitude across echoes, $\overline{\TE}$ is the mean echo time, and $\widehat R^*_2$ (in $\text{s}^{-1}$) is the fitted mono-exponential decay rate. Note that $y(t)$ is a derived observation rather than raw k-space data, as its errors include those propagated from the QSM reconstruction, deviate from a complex Gaussian distribution, and may be spatially correlated across voxels.

\subsection{Discrete problem}
Sampling each pool of~\eqref{eq:signalCanon} at the acquired echo times gives the basis vectors
\begin{equation}\label{eq:basis}
\mathbf{b}^\pm=\begin{bmatrix}e^{(-R+u^\pm\chi^\pm)\TE_1}\\ \vdots\\ e^{(-R+u^\pm\chi^\pm)\TE_{N_e}}\end{bmatrix}=\exp\!\big((-R+u^\pm\chi^\pm)\,\mathbf{t}\big),\qquad \mathbf{b}^0=\exp(-R\,\mathbf{t}).
\end{equation}
Collecting the unknowns in $\btheta=(\mathbf{c},\bchi,R)$, the discrete model signal vector $\mathbf{s}(\btheta)=[S(\TE_e)]_e$ is
\begin{equation}\label{eq:model}
\mathbf{s}(\btheta)=c^+\mathbf{b}^++c^-\mathbf{b}^-+c^0\mathbf{b}^0=\mathbf{B}\mathbf{c},\qquad \mathbf{B}=[\mathbf{b}^+\ \mathbf{b}^-\ \mathbf{b}^0]\in\C^{N_e\times3},
\end{equation}
defined on the feasible parameter set
\begin{equation}\label{eq:Theta}
\Theta=\big\{(\mathbf{c},\bchi,R):\ \mathbf{c}\ge\mathbf{0},\ \1^\top\mathbf{c}=1,\ 0\le\chi^+\le\chi_{\max},\ -\chi_{\max}\le\chi^-\le0,\ 0\le R\le R_{\max}\big\}.
\end{equation}
By~\eqref{eq:signalCanon2}, $\mathbf{s}=e^{-R\mathbf{t}}\odot\bbeta$, where $\bbeta=c^+e^{u^+\chi^+\mathbf{t}}+c^-e^{u^-\chi^-\mathbf{t}}+c^0\1$ is the signal mixture with the common relaxation decay factored out. Because the shared factor $e^{-R\mathbf{t}}$ is real and positive, it affects only the magnitude, so all phase information is carried by $\bbeta$, and the observations are the samples of~\eqref{eq:signalSynth}, $\mathbf{y}=[y(\TE_e)]_e$.

Model~\eqref{eq:model} is linear in $\mathbf{c}$ once $(\bchi,R)$ are fixed, a separable structure well known in nonlinear least squares~\cite{golub1973differentiation}. Geometrically, because $\mathbf{c}$ lies in the probability simplex, $\mathbf{s}$ lies in the triangle in $\C^{N_e}$ whose vertices are the three pool signals $\mathbf{b}^+$, $\mathbf{b}^-$, and $\mathbf{b}^0$: the susceptibilities and $R$ position the vertices, while the concentrations select a point inside the triangle.

Across a set of voxels $V$, the observations form a map $\mathbf{y}:V\to\C^{N_e}$. Since~\eqref{eq:model} contains no spatial coupling between voxels (which has already been resolved by the per-echo dipole inversion), the whole-brain inverse problem decouples into $|V|$ independent voxel problems, enabling massively parallel voxel-wise computation.

\subsection{Output maps}\label{sec:maps}
The products $c^\pm\chi^\pm$ alone do not capture the nonlinear phase and magnitude evolution of the multi-pool signal. Chen \textit{et al.}~\cite{chen2021decompose} therefore define effective component susceptibilities from the phase of the fitted sub-voxel signals. We first specify how the phase is tracked across echoes, and then define the output maps.

Because the principal argument $\arg$ is defined only modulo $2\pi$, the wrapping operator maps any real phase to its principal value in $(-\pi,\pi]$,
\begin{equation}\label{eq:wrap}
\wrap(x)=x-2\pi\Big\lceil\frac{x-\pi}{2\pi}\Big\rceil,
\end{equation}
so that $\arg z=\wrap(\vartheta)$ for any phase $\vartheta$ of $z\neq0$, and $\arg(z_1\bar z_2)=\wrap(\arg z_1-\arg z_2)$. For a discrete signal $\mathbf{m}=(m_1,\dots,m_{N_e})\in\C^{N_e}$ with nonzero entries, we define its \emph{unwrapped phase} $\mathcal{U}(\arg\mathbf{m})\in\R^{N_e}$ as the phase tracked continuously from $t=0$: starting from phase $0$ ($m_0=1$) at $\TE_0=0$, it accumulates the wrapped phase increments between consecutive echoes~\cite{itoh1982analysis}. The effective susceptibility of $\mathbf{m}$ is then defined as the ratio of the summed unwrapped phases to $p$ times the summed echo times:
\begin{equation}\label{eq:slope}
\Psi(\mathbf{m})=\frac{\sum_{e=1}^{N_e}\mathcal{U}(\arg\mathbf{m})_e}{p\sum_{e=1}^{N_e}\TE_e}.
\end{equation}
For a single-pool signal with linear phase $p\chi t$, $\Psi(\mathbf{m})$ returns $\chi$ exactly.

The paramagnetic and diamagnetic component susceptibilities (PCS and DCS) are defined as the effective susceptibilities of a voxel in which the opposing source is replaced by the neutral medium, and the composite susceptibility $\chi_{\mathrm{comp}}$ is that of the full three-pool signal:
\begin{align}
\mathrm{PCS}&=\Psi\big(c^+\mathbf{b}^++(c^0+c^-)\,\mathbf{b}^0\big),\label{eq:pcs}\\
\mathrm{DCS}&=\Psi\big(c^-\mathbf{b}^-+(c^0+c^+)\,\mathbf{b}^0\big),\label{eq:dcs}\\
\chi_{\mathrm{comp}}&=\Psi(\mathbf{s}).\label{eq:comp}
\end{align}
Each of these vectors has the form $e^{-R\mathbf{t}}\odot\mathbf{g}$ with $\mathbf{g}$ independent of $R$, since a positive real multiplier does not alter the complex argument, PCS, DCS, and $\chi_{\mathrm{comp}}$ are independent of $R$. Expanding each signal at short echo times (Theorem~\ref{thm:echo} in Appendix~\ref{app:echo}) shows that as $\TE_e\to0$, where $\mathcal{U}$ acts as the identity, PCS, DCS, and $\chi_{\mathrm{comp}}$ converge to $c^+\chi^+$, $c^-\chi^-$, and the net susceptibility $c^+\chi^++c^-\chi^-$, respectively. At practical echo times, however, they differ substantially because the active pool decays faster than the neutral pool. For example, with $\mathbf{c}=(0.3,0.3,0.4)$ and $\bchi=(0.15,-0.10)$\,ppm at 3\,T, $\mathrm{PCS}=0.012$\,ppm whereas $c^+\chi^+=0.045$\,ppm. This difference reflects the nonlinear phase evolution that the component maps are designed to capture~\cite{chen2021decompose}.

\subsection{Geometrical View}\label{sec:echo}

\begin{figure}
\centering
\includegraphics[width=1\linewidth]{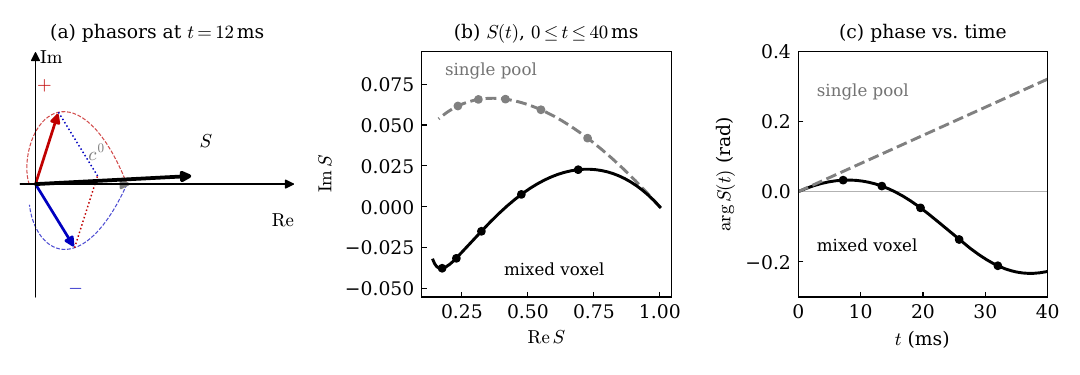}
\caption{Phasor view of the three-pool model~\cite{chen2021decompose} (Eq.~\eqref{eq:signalCanon}) at 3\,T for $\mathbf{c}=(0.3,0.3,0.4)$, $\bchi=(0.15,-0.10)$,ppm, and $R=20$,s$^{-1}$. (a)~At $t=12$\,ms, the paramagnetic (red) and diamagnetic (blue) phasors rotate in opposite directions and nearly cancel in the sum $S$ (black), which remains close to the neutral phasor (gray). Dashed curves show the trajectories of the two source phasors, and dotted lines indicate vector addition. (b)~Trajectory of $S(t)$ for $0\le t\le40$\,ms (black) compared with a single-pool signal sharing the same first-order terms (gray, dashed), and dots mark the five echoes of the \textit{in vivo} protocol ($\TE = 7.2\text{-}32.0$,ms). (c)~Phase of the two signals versus $t$: both start with the same initial slope, determined by the net susceptibility, but the phase of the mixed voxel bends and changes sign as the faster-decaying paramagnetic pool attenuates, leaving the diamagnetic pool dominant at later echoes.}
\label{fig:phasor}
\end{figure}

The model in~\eqref{eq:signalCanon} admits an intuitive geometric interpretation. Each term in~\eqref{eq:signalCanon} represents a rotating phasor in the complex plane: the paramagnetic phasor rotates counter-clockwise, the diamagnetic phasor rotates clockwise, and the neutral phasor remains stationary on the real axis. In addition, the two source phasors shrink faster than the neutral phasor. The composite signal $S(t)$ is the vector sum of these three arrows (Fig.~\ref{fig:phasor}a).

At early echoes, the two opposing rotations nearly cancel, so $S(t)$ stays close to the neutral phasor and its phase remains near zero. At later echoes, the faster-decaying source phasor fades, breaking the cancellation and causing the phase trajectory of $S(t)$ to bend (Fig.~\ref{fig:phasor}b,c). This curvature provides the information needed to separate the sources. The formal statements and proofs are given in Appendix~\ref{app:echo} (Theorem~\ref{thm:echo} and Corollary~\ref{cor:app}).

At short echo times, $\arg S(t)$ grows linearly with slope proportional to the net susceptibility $c^+\chi^++c^-\chi^-$, while $\log|S(t)|$ decays linearly at the rate $R+a\,(c^+\chi^++c^-|\chi^-|)$. Standard linear-phase QSM and mono-exponential $R_2^*$ fitting measure only these two first-order quantities, so a voxel containing high concentrations of both iron and myelin and a voxel containing neither can produce identical first-order measurements.

In a mixed voxel, however, the apparent susceptibility $\chi_{\mathrm{app}}(t)=\arg S(t)/(p\,t)$—which is what a QSM map reconstructed from the echo at time $t$ estimates—is not constant. To first order in $t$, it drifts linearly with echo time (Corollary~\ref{cor:app}):
\begin{equation}\label{eq:driftMain}
\chi_{\mathrm{app}}(t)\approx c^+\chi^++c^-\chi^--a\,\kappa\,t,\qquad \kappa=c^+(1-c^+)(\chi^+)^2-c^-(1-c^-)(\chi^-)^2 .
\end{equation}
The sign of the slope $-a\kappa$ indicates which source dominates the quadratic phase drift. For the mixed voxel in Fig.~\ref{fig:phasor}, $\chi_{\mathrm{app}}(t)$ shifts from $0.0085$\,ppm at $7.2$\,ms to $-0.0123$\,ppm at $32$\,ms, closely matching the values of $0.0089$\,ppm and $-0.0122$\,ppm predicted by~\eqref{eq:driftMain}. Whereas an echo-combined QSM averages out this temporal drift, reconstructing QSM per echo preserves it—explaining why the synthesized signal~\eqref{eq:signalSynth} must be constructed from per-echo QSM maps.

\section{Methods}\label{sec:methods}
\subsection{CANON-QSM}\label{sec:theoryCanonQSM}

Following~\cite{chen2021decompose}, the parameters are estimated by alternating between three blocks, each minimized with the other two fixed:
\begin{align}
\text{(A)}\quad & \underset{\mathbf{c}\ge\mathbf{0},\ \1^\top\mathbf{c}=1}{\text{minimize}}\ \norm{\mathbf{B}\mathbf{c}-\mathbf{y}}^2, \label{eq:subA}\\
\text{(B)}\quad &\underset{0\le R\le R_{\max}}{\text{minimize}}\ \norm{\log\mathbf{s}(R)-\log\mathbf{y}}^2, \label{eq:subB}\\
\text{(C)}\quad & \underset{\substack{0\le\chi^+\le\chi_{\max} \\ -\chi_{\max}\le\chi^-\le0}}{\text{minimize}}\ \norm{\log\mathbf{s}(\bchi)-\log\mathbf{y}}^2. \label{eq:subC}
\end{align}
Block (A) uses a least-squares cost on the signal, in which the model is linear in $\mathbf{c}$. Blocks (B) and (C) use a cost on the complex logarithm, which separates log-magnitude and phase. The following subsections describe how CANON-QSM solves each block. The derivations are given in Appendix~\ref{app:blocks}.

\subsubsection{Subproblem (A): Concentrations in Closed Form}\label{sec:subproblemA}
To prevent a pool from vanishing early in the fit, we impose a lower bound (floor) $c_f\in[0,1/3)$ on every concentration by writing $\mathbf{c}=c_f\1+(1-3c_f)\,\mathbf{q}$, where $\mathbf{q}=(q^+,q^-,q^0)$ is nonnegative and sums to one. With $\mathbf{x}=(q^+,q^-)^\top$ and $q^0=1-q^+-q^-$, define the floor-corrected data $\tilde{\mathbf{y}}=(\mathbf{y}-c_f\mathbf{B}\1)/(1-3c_f)$, the source-versus-neutral contrasts $\mathbf{d}^\pm=\mathbf{b}^\pm-\mathbf{b}^0$, $\mathbf{D}=[\mathbf{d}^+\ \mathbf{d}^-]$, $\mathbf{z}=\tilde{\mathbf{y}}-\mathbf{b}^0$, and
\begin{equation}\label{eq:Gh}
\mathbf{G}=\Re(\mathbf{D}^\Hm\mathbf{D})\in\R^{2\times2},\qquad \mathbf{h}=\Re(\mathbf{D}^\Hm\mathbf{z})\in\R^2.
\end{equation}

\begin{theorem}[Exact solution of subproblem (A)]\label{thm:blockA}
Block (A) with the floor is equivalent to the two-variable convex quadratic program
\begin{equation}\label{eq:qp}
\underset{\mathbf{x}\in\Tri}{\text{minimize}}\ \tfrac12\mathbf{x}^\top\mathbf{G}\mathbf{x}-\mathbf{h}^\top\mathbf{x},\qquad \Tri=\{\mathbf{x}\ge\mathbf{0},\ x^++x^-\le1\}.
\end{equation}
Its global minimizer is the lowest-cost point among four candidates: the unconstrained minimizer $\mathbf{G}^{-1}\mathbf{h}$ (if it lies inside $\Tri$), and the minimizer on each of the three edges of $\Tri$,
\begin{equation}\label{eq:edges}
\big(0,\ \clip_{[0,1]}\tfrac{h_2}{G_{22}}\big),\quad
\big(\clip_{[0,1]}\tfrac{h_1}{G_{11}},\ 0\big),\quad
\big(\tau,\ 1-\tau\big),\ \tau=\clip_{[0,1]}\tfrac{\Re(\mathbf{e}^\Hm\mathbf{w})}{\norm{\mathbf{e}}^2},
\end{equation}
with $\mathbf{e}=\mathbf{d}^+-\mathbf{d}^-$ and $\mathbf{w}=\mathbf{z}-\mathbf{d}^-$. This holds also when $\mathbf{G}$ is singular.
\end{theorem}

Geometrically, block (A) finds the point of the triangle with vertices $\mathbf{b}^+$, $\mathbf{b}^-$, $\mathbf{b}^0$ that is closest to the data. $\mathbf{G}$ is the real Gram matrix of the two triangle edges that leave $\mathbf{b}^0$. The closest point is either the orthogonal projection onto the plane of the triangle, if it falls inside, or lies on one of the three edges. The edges correspond to one pool being absent ($c^+$, $c^-$ or $c^0$ at the floor), and the vertices of $\Tri$ (a single pool) appear when a ratio is clipped. The update is closed-form and needs no iterations.

\subsubsection{Subproblem (B): Baseline Relaxation in Closed Form}\label{sec:subproblemB}
Since $\mathbf{s}=e^{-R\mathbf{t}}\odot\bbeta$ and $e^{-R\mathbf{t}}$ is real and positive, $R$ affects only the log-magnitude of the model and not its phase.

\begin{proposition}[Exact solution of block (B)]\label{prop:blockB}
Let $\brho=\log|\mathbf{y}|-\log|\bbeta|$. The cost in~\eqref{eq:subB} equals $\norm{R\,\mathbf{t}+\brho}^2$ plus a term that does not depend on $R$, and its minimizer over $[0,R_{\max}]$ is
\begin{equation}\label{eq:blockB_sol}
R^\star=\clip_{[0,R_{\max}]}\!\Big(-\frac{\mathbf{t}^\top\brho}{\mathbf{t}^\top\mathbf{t}}\Big).
\end{equation}
\end{proposition}

This is a log-linear $R_2^*$ fit, through the origin, of $|\mathbf{y}|/|\bbeta|$: the data with the decay explained by the two sources divided out.
\subsubsection{Subproblem (C): Susceptibilities by Levenberg-Marquardt}\label{sec:subproblemC}

The phase part of~\eqref{eq:subC}, $\arg\mathbf{s}-\arg\mathbf{y}$, jumps by $2\pi$ whenever one of the two phases wraps and the other does not. This occurs for large venous susceptibilities, at high field and at late echoes, and it adds spurious penalties of order $(2\pi)^2$ to the cost. We therefore use the residual
\begin{equation}\label{eq:r}
\mathbf{r}(\bchi)=\begin{bmatrix}\log|\mathbf{s}|-\log|\mathbf{y}|\\ \arg(\mathbf{s}\odot\bar{\mathbf{y}})\end{bmatrix}\in\R^{2N_e},
\end{equation}
whose phase part is the phase difference wrapped to $(-\pi,\pi]$, $\arg(\mathbf{s}\odot\bar{\mathbf{y}})=\wrap(\arg\mathbf{s}-\arg\mathbf{y})$ with $\wrap$ as in~\eqref{eq:wrap}. In other words, $\norm{\mathbf{r}}^2$ is the original cost of~\eqref{eq:subC} with every spurious $2\pi$ jump removed, and it coincides with the original cost whenever no echo has such a jump. Block (C) thus becomes a bounded nonlinear least-squares problem in two variables.

With $\mathbf{s}^\pm=c^\pm\mathbf{b}^\pm$, the Jacobian of $\mathbf{r}$ has columns
\begin{equation}\label{eq:jac}
\mathbf{j}^\pm=\begin{bmatrix}\Re\,\mathbf{v}^\pm\\ \Im\,\mathbf{v}^\pm\end{bmatrix},\qquad
\mathbf{v}^\pm=u^\pm\,(\mathbf{s}^\pm\oslash\mathbf{s})\odot\mathbf{t}.
\end{equation}
The sensitivity to $\chi^\pm$ is thus the fraction of the signal coming from pool $\pm$, multiplied by its complex coefficient and by $\TE$: late echoes and abundant pools are the most informative. Each Levenberg-Marquardt (LM) step~\cite{levenberg1944method,marquardt1963algorithm,more1978levenberg} solves
\begin{equation}\label{eq:lm}
\big(\mathbf{J}^\top\mathbf{J}+\lambda\bar\sigma\mathbf{I}\big)\bdelta=-\mathbf{J}^\top\mathbf{r},\qquad \bar\sigma=\tfrac12\tr(\mathbf{J}^\top\mathbf{J}),
\end{equation}
a $2\times2$ system solved in closed form by Cramer's rule. The step interpolates between a Gauss-Newton step ($\lambda\to0$) and a short gradient-descent step ($\lambda\to\infty$). Scaling the damping by $\bar\sigma$ makes $\lambda$ independent of the units of $\chi$. If a concentration reaches zero, then $\mathbf{s}^\pm=\mathbf{0}$, the corresponding Jacobian column vanishes, and $\chi^\pm$ can no longer be updated. The floor $c_f$ prevents this:

\begin{proposition}\label{prop:lm}
If $c^\pm\ge c_f>0$, $\TE_e>0$ for all $e$, and $\mathbf{s}$ has nonzero entries, then $\mathbf{j}^\pm\neq\mathbf{0}$ and the matrix in~\eqref{eq:lm} is positive definite for every $\lambda>0$. Whenever $\mathbf{J}^\top\mathbf{r}\neq\mathbf{0}$, the LM step is a strict descent direction for $\norm{\mathbf{r}}^2$.
\end{proposition}

We therefore start with a large floor and decrease it geometrically over the iterations, so that both sources remain observable while the fit is far from the solution, and the floor no longer biases the final estimate.

Exact unwrapping of the output maps requires $|p\chi|\max_e(\TE_e-\TE_{e-1})<\pi$ with $\TE_0=0$~\cite{itoh1982analysis}, which for a single pool means $|\chi|<0.82$\,ppm at 3\,T and $|\chi|<0.31$\,ppm at 7\,T for our protocols. The bound $\chi_{\max}$ should therefore be of the same order as these values. Chen \textit{et al.} used $\chi_{\max}=0.5$\,ppm~\cite{chen2021decompose}. In our \textit{in silico} experiments we set $\chi_{\max}=\max(0.2,\,1.5\max|\chi_{\mathrm{GT}}|)=0.48$\,ppm, where $\chi_{\mathrm{GT}}$ is the ground-truth susceptibility of the phantom. For \textit{in vivo} data $\chi_{\max}$ is chosen per dataset, since no ground truth is available.

\subsection{Multi-start strategy and output selection}\label{sec:global}
Block (C) is non-convex, so each voxel runs $K$ initializations (restarts) in parallel. The deterministic restart starts from $\bchi=(0.1\chi_{\max},-0.1\chi_{\max})$ and from the two-point estimate $R=-\log(|y_{N_e}|/|y_1|)/(\TE_{N_e}-\TE_1)$. The other $K-1$ restarts draw $\chi^+/\chi_{\max}$ and $-\chi^-/\chi_{\max}$ uniformly from $[0.01,0.99]$ and $R$ uniformly from $[0,\max\{60,2\bar R_0\}]$\,s$^{-1}$, where $\bar R_0$ is the mean two-point estimate. Block (A) is solved first, so the initial concentrations play no role.

Blocks (B) and (C) reduce the log-domain cost, whereas block (A) reduces the signal-domain cost, so no single cost decreases monotonically over the alternation. After each outer iteration, we therefore select the restart with the lowest score
\begin{equation}\label{eq:score}
\mathcal{S}=\frac{\norm{\mathbf{s}-\mathbf{y}}^2}{\norm{\mathbf{y}}^2}+\frac{\norm{\mathbf{r}}^2}{N_e},
\end{equation}
which weights both costs, and the output is the best state over all iterations.

\subsection{Implementation}
Algorithm~\ref{alg:canon} summarizes the solver. Because all voxels and restarts are mutually independent, we process batches of voxels $\times\,K$ restarts in parallel on a GPU in PyTorch~\cite{paszke2019pytorch}, using spatial chunks to bound memory consumption. The computational complexity per volume is $\mathcal{O}\big(T\,K\,(n_{\mathrm{LM}})\,N_e\,N_{\mathrm{vox}}\big)$, which scales linearly with the number of voxels and echoes.

\subsection{Experimental Design}
\label{sec:methodsExperimentalDesign}

We evaluated CANON-QSM in the following experiments:

\textbf{In silico:} We used the realistic susceptibility-separation brain phantom~\cite{ridani2026realistic}, which provides ground-truth isotropic $\chi^{+}$ and $\chi^{-}$ maps (matrix size $256\times 320 \times 320$, $0.64$\,mm isotropic resolution, $\mathrm{FOV} = 164\times 205 \times205$\,mm$^3$). Five echoes ($\TE=4\text{-}36$\,ms, $\Delta\TE=8$\,ms) were simulated at $7$\,T. Complex Gaussian noise with standard deviation $\sigma = \text{mean}(M_{1})/\text{SNR}$, where $M_{1}$ is the first-echo magnitude, was added directly to the synthesized per-echo input signal to isolate solver performance from upstream preprocessing errors. We also evaluated ROI-specific reconstruction error and multi-start convergence reliability at 3\,T and 7\,T.

\textbf{In vivo:} One healthy volunteer was scanned at $3$\,T using a 3D multi-echo GRE sequence with institutional ethics approval (voxel size $0.60\times0.60\times1.0$\,mm$^3$, $\mathrm{TR}/\TE_1/\Delta\TE = 43.4/7.2/6.2$\,ms, 5 echoes, bandwidth $550.5$\,Hz/pixel, matrix size $352\times352\times160$, flip angle $17^\circ$). Preprocessing comprised ROMEO phase unwrapping~\cite{dymerska2021phase}, PDF background field removal~\cite{liu2011novel}, WH-QSM harmonic remnant estimation~\cite{milovic2019weak}, and per-echo dipole inversion via DEMIX-QSM~\cite{contreras2026demixing}.

In both experiments, CANON-QSM was compared against DECOMPOSE-QSM run for 10 iterations as recommended in~\cite{chen2021decompose}, using the reference implementation in QSM-CI~\cite{stewart2024qsmci}. We report composite, PCS, and DCS maps, and evaluate reconstruction accuracy using NRMSE and XSIM~\cite{milovic2025xsim}.

\section{Results}\label{sec:results}

\begin{figure}
\centering
\includegraphics[width=1\linewidth]{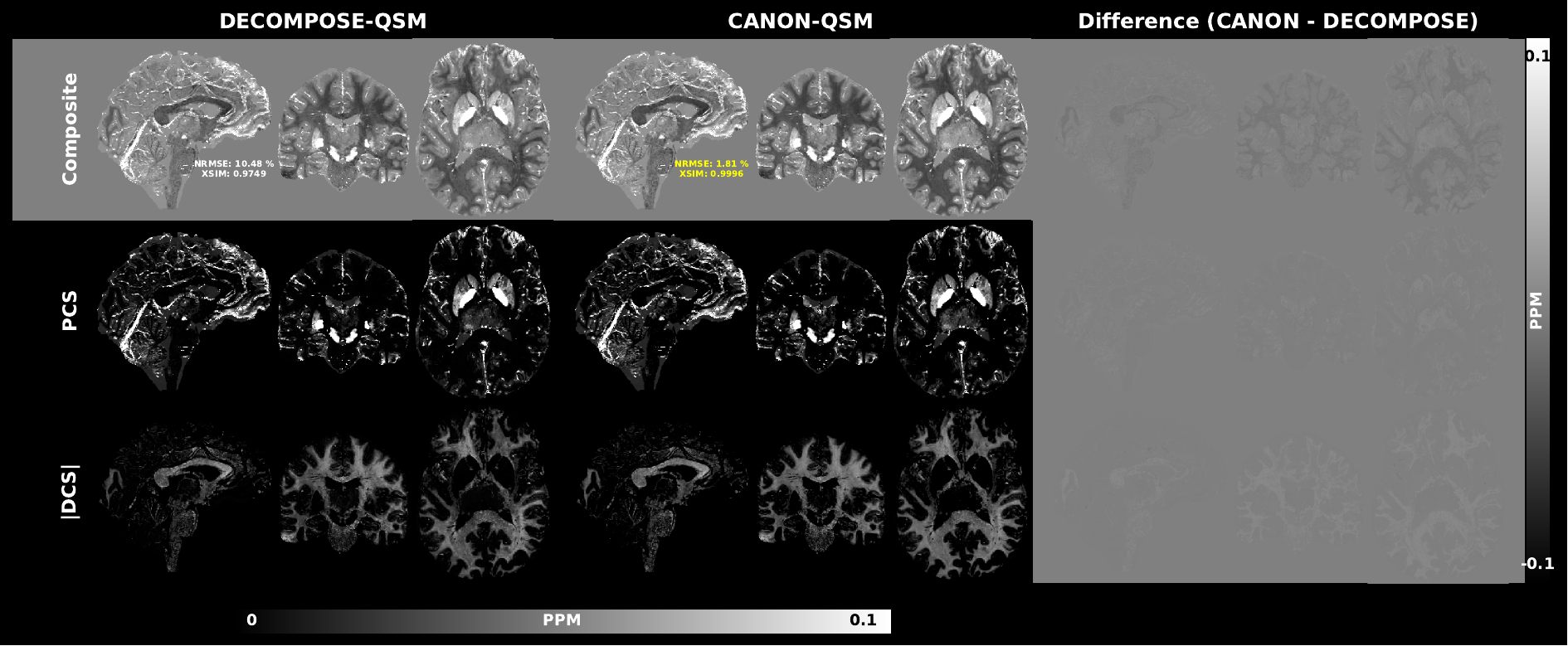}
\caption{\textit{In silico} 7\,T phantom comparison (noiseless) in sagittal, coronal, and axial views. Both CANON-QSM ($T=20$, $K=20$) and DECOMPOSE-QSM (10 iterations) resolve the three-pool model across the Composite, paramagnetic component susceptibility (PCS), and magnitude diamagnetic component susceptibility ($\vert{}\mathrm{DCS}\vert{}$) maps. CANON-QSM yields sharper tissue contrast, smaller inter-method differences, and lower composite reconstruction error (normalized root-mean-square error [NRMSE]: $1.81\%$ vs.\ $10.48\%$, XSIM: $0.9996$ vs.\ $0.9749$).}
\label{fig:fig01}
\end{figure}

\begin{figure}
\centering
\includegraphics[width=1\linewidth]{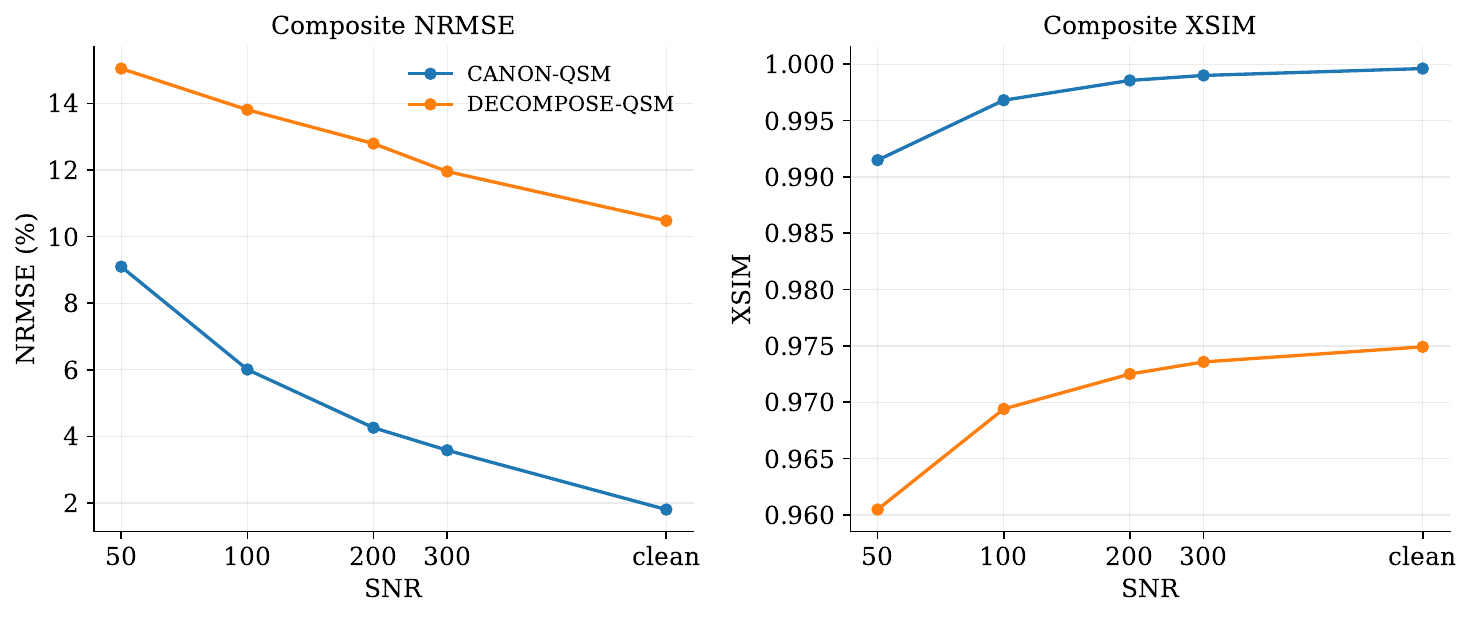}
\caption{Noise robustness on the \textit{in silico} 7\,T brain phantom. CANON-QSM ($T=20$, $K=20$) consistently achieves lower composite normalized root-mean-square error (NRMSE) and higher structural similarity (XSIM) than DECOMPOSE-QSM (10 iterations) across all signal-to-noise ratio (SNR) levels ($\mathrm{SNR}=50\text{-}300$ and noiseless [clean]). Although reconstruction accuracy for both methods degrades with increasing noise, CANON-QSM maintains superior performance across the entire SNR range.}
\label{fig:fig02}
\end{figure}

\begin{figure}
\centering
\includegraphics[width=1\linewidth]{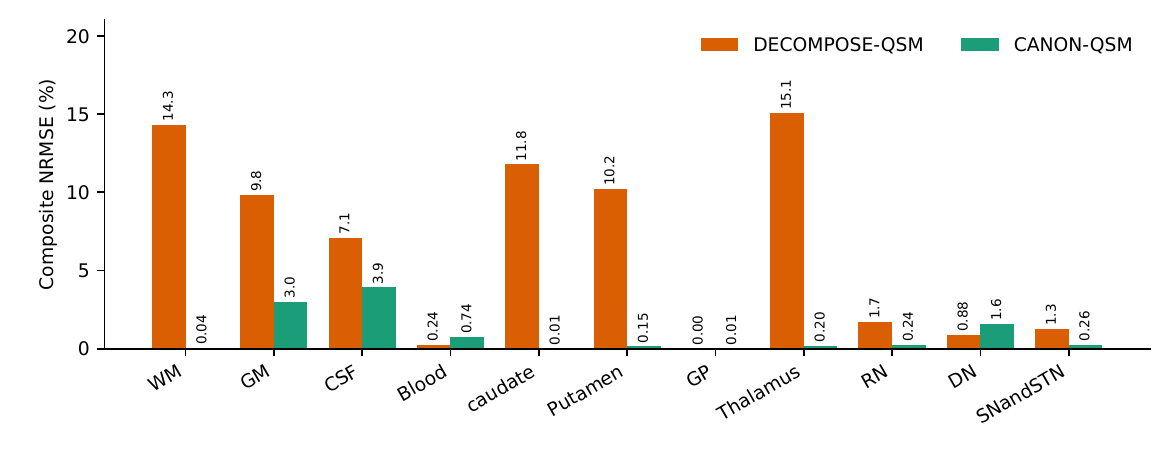}
\caption{Composite normalized root-mean-square error (NRMSE) in 11 segmented structures of the noiseless \textit{in silico} 7\,T phantom. DECOMPOSE-QSM was run for 10 iterations and CANON-QSM with $T = 20$ and $K = 20$. CANON-QSM achieves lower error in 8 of 11 structures, whereas DECOMPOSE-QSM yields lower error in venous blood and the dentate nucleus (DN), and both methods achieve negligible error in the globus pallidus (GP). Abbreviations: CSF, cerebrospinal fluid, GM, gray matter, RN, red nucleus, SNandSTN, substantia nigra and subthalamic nucleus, WM, white matter.}
\label{fig:fig03}
\end{figure}

\textbf{In silico:} Fig.~\ref{fig:fig01} compares the noiseless reconstructions. Both methods separate the three pools across the composite, PCS, and $|\mathrm{DCS}|$ maps, with CANON-QSM preserving sharper tissue contrast and exhibiting smaller residual errors relative to the ground truth. Across the whole-brain volume, CANON-QSM ($T=20$, $K=20$) achieved a lower composite NRMSE than DECOMPOSE-QSM (10 iterations) both in the noiseless case ($1.8\%$ vs.\ $10.5\%$) and at $\mathrm{SNR}=100$ ($6.0\%$ vs.\ $13.8\%$), as well as consistently higher XSIM across all noise levels (Fig.~\ref{fig:fig02}). In the noiseless ROI analysis (Fig.~\ref{fig:fig03}), CANON-QSM yielded lower NRMSE in 8 of 11 segmented structures, with the largest improvements in white matter ($0.04\%$ vs.\ $14.3\%$), caudate ($0.01\%$ vs.\ $11.8\%$), thalamus ($0.20\%$ vs.\ $15.1\%$), and putamen ($0.15\%$ vs.\ $10.2\%$). DECOMPOSE-QSM yielded slightly lower NRMSE in venous blood ($0.24\%$ vs.\ $0.74\%$) and the dentate nucleus ($0.88\%$ vs.\ $1.6\%$), while both methods achieved negligible error in the globus pallidus.

\begin{figure}
\centering
\includegraphics[width=1\linewidth]{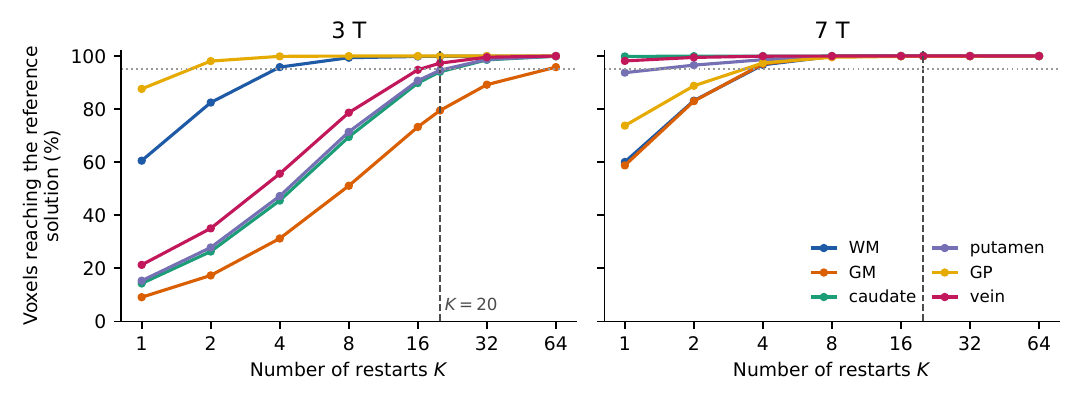}
\caption{Reliability of the multi-start strategy on the \textit{in silico} phantom at $\mathrm{SNR}=100$ across six tissue classes. Curves show the percentage of voxels in which the selected solution recovers paramagnetic and diamagnetic component susceptibilities (PCS and DCS) within $0.005$,ppm of the reference solution (defined as the best-scoring solution among 512 random starts plus a start at the ground-truth parameters) as a function of the number of restarts $K$ at 3\,T (left) and 7\,T (right). The vertical dashed line marks $K=20$ (the default value used in this work), and the horizontal dotted line marks the $95\%$ convergence threshold. Abbreviations: GM, gray matter, GP, globus pallidus, WM, white matter.}
\label{fig:fig04}
\end{figure}

Fig.~\ref{fig:fig04} shows the percentage of voxels in which the selected solution converges to within $0.005$\,ppm of the 512-start reference solution as a function of the number of restarts $K$. At 7\,T, every tissue reached the reference solution in at least $99.9\%$ of voxels for $K\ge16$. At 3\,T, convergence was slower: with $K=20$, the convergence rate reached $79.5\%$ in cortical gray matter, $94\%$ in the caudate, $95\%$ in the putamen, $97\%$ in veins, and nearly $100\%$ in white matter and the globus pallidus, increasing to $K=64$ raised convergence to at least $96\%$ across all tissues. Thus, $K=20$ is sufficient at 7\,T and in most structures at 3\,T, whereas cortical gray matter at 3\,T benefits from additional restarts.

\begin{figure}
\centering
\includegraphics[width=1\linewidth]{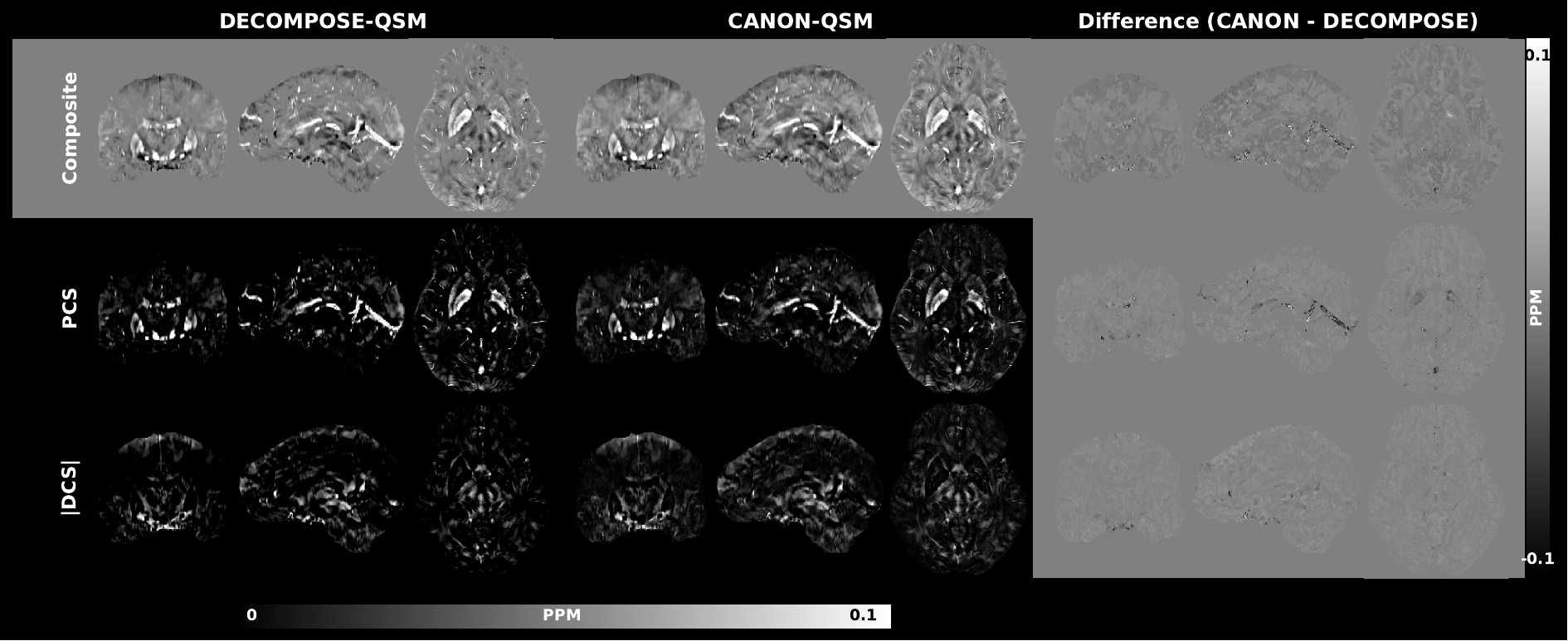}
\caption{\textit{In vivo} 3\,T whole-brain results in coronal, sagittal, and axial views. CANON-QSM ($T=20$, $K=20$) yields consistent susceptibility separation across the Composite, paramagnetic component susceptibility (PCS), and magnitude diamagnetic component susceptibility ($\vert{}\mathrm{DCS}\vert{}$) maps. In contrast, DECOMPOSE-QSM (at the recommended 10 iterations) displays severe fitting artifacts in the great vein of Galen (sagittal view) in both the composite and PCS maps, falsely assigning paramagnetic venous voxels to the diamagnetic pool (consistent with~\cite{fuchs2024practical}), alongside reduced tissue contrast.}
\label{fig:fig05}
\end{figure}

\textbf{In vivo:} Fig.~\ref{fig:fig05} displays the \textit{in vivo} reconstructions. DECOMPOSE-QSM produced severe fitting artifacts in the great vein of Galen (sagittal view) in both the composite and PCS maps—erroneously assigning paramagnetic venous voxels to the diamagnetic pool, consistent with~\cite{fuchs2024practical}—and exhibited reduced tissue contrast at 10 iterations. CANON-QSM eliminated these venous artifacts and yielded clear delineation of deep gray matter nuclei and white matter tracts. For the $352\times352\times160$ volume (Table~\ref{tab:runtime}), CANON-QSM completed in $113$\,s using $1.68$\,GB of peak GPU memory, compared with approximately $5$\,h for the MATLAB reference implementation of DECOMPOSE-QSM, corresponding to a $\approx160\times$ speedup.

\begin{table}[t]
\caption{Whole-brain runtime (\textit{in vivo}, $352\times352\times160$, 5 echoes).}\label{tab:runtime}
\centering\small
\begin{tabular}{@{}llll@{}}
\toprule
Method & Hardware & Time & Peak memory\\
\midrule
DECOMPOSE-QSM (Matlab Reference) & CPU, 18 threads & $\approx5$\,h & --\\
CANON-QSM ($T=20$, $K=20$) & GPU (RTX 5090) & $113$\,s & $1.68$\,GB (GPU)\\
\bottomrule
\end{tabular}
\end{table}


\section{Discussion}
\label{sec:discussion}

Susceptibility source separation is a rapidly growing area of QSM, motivated by the fact that the bulk susceptibility value can remain near zero even when paramagnetic and diamagnetic sources are both abundant within a voxel. In this work, we introduced CANON-QSM, a solver for the three-pool model of DECOMPOSE-QSM~\cite{chen2021decompose} in which two of the three alternating blocks are solved in closed form and the third is formulated as a two-variable Levenberg-Marquardt problem with a residual that is insensitive to phase wrapping. We also analyzed the physics of the echo train. Expanding the signal in powers of echo time shows that linear-phase QSM and mono-exponential $R_2^*$ fitting measure only the net susceptibility and the total decay, whereas the sources are separated by the second-order drift of the apparent susceptibility with echo time. On the \textit{in silico} phantom, CANON-QSM reached a lower composite NRMSE than DECOMPOSE-QSM (with 10 iterations) without noise ($1.8\%$ vs.\ $10.5\%$), and it processed a whole-brain volume in under two minutes on a GPU. In \textit{in vivo} data, the whole-brain reconstruction took $113$\,s, compared with approximately $5$\,h for the reference implementation, and the venous fitting artifacts of the reference~\cite{fuchs2024practical} were eliminated.

The substantial speedup stems from replacing iterative inner-loop solvers with exact block updates, which removes inner iterations and makes parallel restarts per voxel computationally affordable. Combined with the decaying concentration floor, which prevents pool gradients from vanishing early in the fit, and the wrapped phase residual, this multi-start strategy also explains the improved accuracy and the absence of $2\pi$ wrapping artifacts in large veins. Nevertheless, because CANON-QSM shares the physical forward model of DECOMPOSE-QSM, it inherits the assumptions of the static dephasing regime~\cite{yablonskiy1994theory}, isotropic susceptibility, and error propagation from the per-echo QSM inputs. Finally, while our \textit{in vivo} proof of concept demonstrates feasibility at $3$\,T, a larger cohort and histological or clinical dataset will be necessary to establish robustness and generalizability across diverse acquisition protocols and pathological scenarios.

\section{Conclusions}
\label{sec:conclusion}

We introduced CANON-QSM, a solver for the three-pool susceptibility source separation model in which two of the three alternating blocks are solved in closed form and the third is formulated as a two-variable Newton-type (Levenberg-Marquardt) problem with a residual that is robust to phase wrapping. An expansion of the signal in powers of echo time explains why the model requires per-echo QSM reconstructions: linear-phase QSM and mono-exponential $R_2^*$ fitting capture only the net susceptibility and the total decay, whereas the sources are separated by the second-order drift of the apparent susceptibility with echo time, at the cost of information that vanishes rapidly for short echo trains. Because every voxel is independent and every update reduces to elementwise operations, the solver runs voxel-parallel on a GPU, achieving whole-brain reconstructions in under two minutes without venous phase-wrapping artifacts.

Beyond a faster solver for this three-pool model, a new theoretical formulation is needed to accurately account for the presence of multiple paramagnetic and diamagnetic species within a voxel without fixing their number \textit{a priori}, as the exact number of sub-voxel compartments is generally unknown.

\section{Acknowledgments}\label{sec:acknowledgments}
We thank ANID: FONDECYT de Iniciaci\'on 11240279, Fondecyt 1231535 and 1270460.
\section{Data Availability Statement}
The source code from this paper is available on GitHub at \texttt{ignacontreras/CANON-QSM}.







\bibliographystyle{unsrt}
\bibliography{references}
\newpage
\appendix
\section{Theorems and proofs for Section~\ref{sec:echo}}\label{app:echo}
We define the following summary quantities:
\begin{align}
\chi_{\mathrm{net}}&=c^+\chi^++c^-\chi^- &&\text{(net susceptibility)},\label{eq:net}\\
\chi_{\mathrm{tot}}&=c^+\chi^++c^-|\chi^-| &&\text{(total susceptibility)},\label{eq:tot}\\
q&=c^+(\chi^+)^2+c^-(\chi^-)^2,\label{eq:q}\\
v_{\mathrm{net}}&=q-\chi_{\mathrm{net}}^2,\qquad
v_{\mathrm{tot}}=q-\chi_{\mathrm{tot}}^2,\label{eq:v}\\
\kappa&=c^+(1-c^+)(\chi^+)^2-c^-(1-c^-)(\chi^-)^2 .\label{eq:kappa}
\end{align}
Here, $\chi_{\mathrm{net}}$ represents the bulk susceptibility measured by traditional QSM, whereas $\chi_{\mathrm{tot}}$ governs the overall signal decay rate. The terms $v_{\mathrm{net}}$ and $v_{\mathrm{tot}}$ quantify the dispersion of the two susceptibilities around these values, and $\kappa$ measures the asymmetry between positive and negative source distributions. Note that $\kappa=c^+(\chi^+)^2-c^-(\chi^-)^2-\chi_{\mathrm{tot}}\chi_{\mathrm{net}}$, since $\chi_{\mathrm{tot}}\chi_{\mathrm{net}}=(c^+)^2(\chi^+)^2-(c^-)^2(\chi^-)^2$ (using $|\chi^-|=-\chi^-$). For a single source mixed with the neutral pool ($c^-=0$, $0<c^+<1$, $\chi^+>0$), we have $\kappa=c^+(1-c^+)(\chi^+)^2>0$.

\begin{theorem}[Echo-time expansion]\label{thm:echo}
Let $\arg S(t)$ be the continuous unwrapped phase of $S(t)$, with $\arg S(0)=0$. For $t$ in a neighborhood of $0$,
\begin{align}
\log|S(t)|&=-\big(R+a\,\chi_{\mathrm{tot}}\big)\,t+\tfrac12\big(a^2\,v_{\mathrm{tot}}-p^2\,v_{\mathrm{net}}\big)\,t^2+\mathcal{O}(t^3),\label{eq:mag}\\[2pt]
\arg S(t)&=p\,\chi_{\mathrm{net}}\,t-a\,p\,\kappa\,t^2+\mathcal{O}(t^3).\label{eq:phase}
\end{align}
\end{theorem}

\begin{corollary}[Apparent susceptibility]\label{cor:app}
The apparent susceptibility, defined as the phase divided by $p\,t$,
\[
\chi_{\mathrm{app}}(t)=\frac{\arg S(t)}{p\,t},
\]
which corresponds to the QSM value reconstructed from the echo acquired at time $t$, satisfies
\begin{equation}\label{eq:app}
\ \chi_{\mathrm{app}}(t)=\chi_{\mathrm{net}}-a\,\kappa\,t+\mathcal{O}(t^2).\ 
\end{equation}
In terms of the original parameters,
\[
\chi_{\mathrm{app}}(t)\approx\big(c^+\chi^++c^-\chi^-\big)-a\Big(c^+(1-c^+)(\chi^+)^2-c^-(1-c^-)(\chi^-)^2\Big)\,t .
\]
\end{corollary}

\begin{proof}[Proof of Theorem~\ref{thm:echo}]
Express the signal as $S(t)=e^{-Rt}\beta(t)$ with
\[
\beta(t)=c^+e^{\omega^+t}+c^-e^{\omega^-t}+c^0,\qquad \omega^+=(-a+ip)\,\chi^+,\quad \omega^-=(a+ip)\,\chi^- ,
\]
where we used $-a|\chi^-|=a\chi^-$. Then $\log S(t)=-Rt+\log\beta(t)$.

Expanding each exponential to second order and using $c^++c^-+c^0=1$ gives
\[
\beta(t)=1+\mu_1\,t+\tfrac12\mu_2\,t^2+\mathcal{O}(t^3),\qquad \mu_1=c^+\omega^++c^-\omega^-,\quad \mu_2=c^+(\omega^+)^2+c^-(\omega^-)^2 .
\]

Since $\beta(0)=1$, $\log\beta(t)$ is analytic near $t=0$ with $\log\beta(0)=0$, applying $\log(1+z)=z-\tfrac12z^2+\mathcal{O}(z^3)$ yields
\[
\log\beta(t)=\mu_1\,t+\tfrac12\big(\mu_2-\mu_1^2\big)\,t^2+\mathcal{O}(t^3).
\]

Evaluating the coefficients, we obtain
\[
\mu_1=c^+(-a+ip)\chi^++c^-(a+ip)\chi^-=-a\,\chi_{\mathrm{tot}}+i\,p\,\chi_{\mathrm{net}} .
\]
Next, $(\omega^+)^2=(a^2-p^2-2iap)(\chi^+)^2$ and $(\omega^-)^2=(a^2-p^2+2iap)(\chi^-)^2$, so with $q$ from~\eqref{eq:q} and $r=c^+(\chi^+)^2-c^-(\chi^-)^2$,
\[
\mu_2=(a^2-p^2)\,q-2iap\,r,\qquad
\mu_1^2=a^2\chi_{\mathrm{tot}}^2-p^2\chi_{\mathrm{net}}^2-2iap\,\chi_{\mathrm{tot}}\chi_{\mathrm{net}} .
\]
Therefore,
\[
\mu_2-\mu_1^2=a^2\big(q-\chi_{\mathrm{tot}}^2\big)-p^2\big(q-\chi_{\mathrm{net}}^2\big)-2iap\,\big(r-\chi_{\mathrm{tot}}\chi_{\mathrm{net}}\big)
=a^2v_{\mathrm{tot}}-p^2v_{\mathrm{net}}-2iap\,\kappa,
\]
where we used the identity $\kappa=r-\chi_{\mathrm{tot}}\chi_{\mathrm{net}}$.

Combining these expressions,
\[
\log S(t)=-\big(R+a\chi_{\mathrm{tot}}\big)t+ip\chi_{\mathrm{net}}t+\tfrac12\big(a^2v_{\mathrm{tot}}-p^2v_{\mathrm{net}}-2iap\,\kappa\big)t^2+\mathcal{O}(t^3).
\]
Taking the real part yields $\log|S(t)|$ in~\eqref{eq:mag}, and the imaginary part yields the unwrapped phase $\arg S(t)$ in~\eqref{eq:phase}. Dividing~\eqref{eq:phase} by $p\,t$ gives~\eqref{eq:app}.  
\end{proof}

\section{Proofs for Section~\ref{sec:theoryCanonQSM}}\label{app:blocks}

\begin{proof}[Proof of Theorem~\ref{thm:blockA}]
Substituting $\mathbf{c}=c_f\1+(1-3c_f)\mathbf{q}$ into the objective gives $\mathbf{B}\mathbf{c}-\mathbf{y}=(1-3c_f)(\mathbf{B}\mathbf{q}-\tilde{\mathbf{y}})$. Since $q^0=1-x^+-x^-$ (with $\mathbf{x}=(x^+,x^-)^\top$), we have $\mathbf{B}\mathbf{q}=\mathbf{b}^0+\mathbf{D}\mathbf{x}$, so $\mathbf{B}\mathbf{q}-\tilde{\mathbf{y}}=\mathbf{D}\mathbf{x}-\mathbf{z}$. For real $\mathbf{x}$,
\[
\norm{\mathbf{D}\mathbf{x}-\mathbf{z}}^2=\mathbf{x}^\top\mathbf{G}\mathbf{x}-2\,\mathbf{h}^\top\mathbf{x}+\norm{\mathbf{z}}^2,
\]
because the imaginary part of the Hermitian matrix $\mathbf{D}^\Hm\mathbf{D}$ is antisymmetric and vanishes in quadratic forms. Hence, minimizing the objective is equivalent to~\eqref{eq:qp}, and $\mathbf{G}$ is positive semidefinite because $\mathbf{x}^\top\mathbf{G}\mathbf{x}=\norm{\mathbf{D}\mathbf{x}}^2$.

On each boundary edge of $\Tri$, the cost reduces to a one-dimensional quadratic function constrained to $[0,1]$. The minimizers over the edges and the unconstrained interior minimizer $\mathbf{G}^{-1}\mathbf{h}$ (when $\mathbf{G}$ is invertible and the point lies in $\Tri$) form the candidate set $\mathcal{C}$. When $\mathbf{G}$ is singular, let $\mathbf{n}\neq\mathbf{0}$ satisfy $\mathbf{G}\mathbf{n}=\mathbf{0}$, then $\mathbf{h}^\top\mathbf{n}=0$ because $\mathbf{D}\mathbf{n}=\mathbf{0}$ implies $\mathbf{h}^\top\mathbf{n}=\Re((\mathbf{D}\mathbf{n})^\Hm\mathbf{z})=0$. The objective is thus constant along lines parallel to $\mathbf{n}$, which intersect the boundary at an edge candidate achieving the same minimum value. Taking the best among all candidates yields the global minimum.
\end{proof}

\begin{proof}[Proof of Proposition~\ref{prop:blockB}]
Since $\mathbf{s}=e^{-R\mathbf{t}}\odot\bbeta$ and $e^{-R\mathbf{t}}$ is real and positive, $\log\mathbf{s}=-R\mathbf{t}+\log|\bbeta|+i\arg\bbeta$. The real part of $\log\mathbf{s}-\log\mathbf{y}$ is $-(R\mathbf{t}+\brho)$, while the imaginary part ($\arg\bbeta-\arg\mathbf{y}$) is independent of $R$. The cost function is thus a convex parabola in $R$, $(\mathbf{t}^\top\mathbf{t})R^2+2(\mathbf{t}^\top\brho)R+\text{const}$, whose unconstrained minimizer is $-\mathbf{t}^\top\brho / \mathbf{t}^\top\mathbf{t}$. Projecting onto $[0,R_{\max}]$ yields~\eqref{eq:blockB_sol}.
\end{proof}

\begin{proof}[Proof of Proposition~\ref{prop:lm}]
Away from branch cuts where the wrapped phase residual satisfies $|\arg(s_e\bar{y}_e)|<\pi$, the partial derivative of $\mathbf{r}$ with respect to $\chi^\pm$ is $\Im(\partial\log\mathbf{s}/\partial\chi^\pm)$, even if $\arg s_e$ crosses $\pm\pi$. From~\eqref{eq:basis}, $\partial\mathbf{b}^\pm/\partial\chi^\pm=u^\pm\mathbf{t}\odot\mathbf{b}^\pm$, yielding $\partial\log\mathbf{s}/\partial\chi^\pm=u^\pm(\mathbf{s}^\pm\oslash\mathbf{s})\odot\mathbf{t}=\mathbf{v}^\pm$ and leading to the Jacobian columns in~\eqref{eq:jac}. If $c^\pm\ge c_f>0$ and $\TE_e>0$, every entry of $\mathbf{v}^\pm$ is nonzero, ensuring $\bar\sigma>0$. The matrix $\mathbf{M}=\mathbf{J}^\top\mathbf{J}+\lambda\bar\sigma\mathbf{I}$ is therefore positive definite for all $\lambda>0$, guaranteeing that $\mathbf{J}^\top\mathbf{r}\neq\mathbf{0}$ defines a strict descent direction for $\norm{\mathbf{r}}^2$.
\end{proof}

\newpage
\section{Algorithm}\label{app:algorithm}

\begin{algorithm}[!h]
\caption{CANON-QSM}\label{alg:canon}
\small
\begin{algorithmic}[1]
\Require $\mathbf{y}$, $\mathbf{t}$, $u^\pm$, $p$, $\chi_{\max}$, $R_{\max}$, $T$, $K$, $n_{\mathrm{LM}}$, floor schedule $\{c_f^{(\ell)}\}_{\ell=1}^T$, $\lambda_0$, $\nu$
\Ensure PCS, DCS, $\chi_{\mathrm{comp}}$, and $(\mathbf{c},R,\bchi)$
\State Initialize $K$ restarts $(\bchi,R)$ (1 deterministic, $K-1$ random), $\lambda\gets\lambda_0$
\For{$\ell=1,\dots,T$}
  \Statex \hspace{\algorithmicindent}\textit{(A) Concentrations}
  \State $c_f\gets c_f^{(\ell)}$,\ \ $\mathbf{b}^\pm\gets\exp\!\big((-R+u^\pm\chi^\pm)\mathbf{t}\big)$,\ $\mathbf{b}^0\gets\exp(-R\mathbf{t})$,\ $\mathbf{d}^\pm\gets\mathbf{b}^\pm-\mathbf{b}^0$
  \State $\mathbf{z}\gets(\mathbf{y}-c_f\mathbf{B}\1)/(1-3c_f)-\mathbf{b}^0$,\ \ $\mathbf{G}\gets\Re(\mathbf{D}^\Hm\mathbf{D})$,\ $\mathbf{h}\gets\Re(\mathbf{D}^\Hm\mathbf{z})$
  \State $\tau\gets\clip_{[0,1]}\big(\Re(\mathbf{e}^\Hm\mathbf{w})/\max(\|\mathbf{e}\|^2, \epsilon)\big)$ with $\mathbf{e}=\mathbf{d}^+-\mathbf{d}^-$, $\mathbf{w}=\mathbf{z}-\mathbf{d}^-$
  \State $\tau_1\gets\clip_{[0,1]}(h_1/G_{11})$ if $G_{11}>0$ else $0$,\ \ $\tau_2\gets\clip_{[0,1]}(h_2/G_{22})$ if $G_{22}>0$ else $0$
  \State $\mathcal{C}\gets\big\{(0,\tau_2),\ (\tau_1,0),\ (\tau,1-\tau)\big\}$,\ \ \textbf{if} $\det\mathbf{G}>0$ and $\mathbf{G}^{-1}\mathbf{h}\in\Tri$ \textbf{then} $\mathcal{C}\gets\mathcal{C}\cup\{\mathbf{G}^{-1}\mathbf{h}\}$
  \State $\mathbf{x}^\star\gets\arg\min_{\mathbf{x}\in\mathcal{C}}\ \norm{\mathbf{D}\mathbf{x}-\mathbf{z}}^2$,\ \ $\mathbf{c}\gets c_f\1+(1-3c_f)\,\big(x^\star_1,\ x^\star_2,\ 1-x^\star_1-x^\star_2\big)^\top$
  \Statex \hspace{\algorithmicindent}\textit{(B) Baseline relaxation}
  \State $\bbeta\gets c^+e^{u^+\chi^+\mathbf{t}}+c^-e^{u^-\chi^-\mathbf{t}}+c^0\1$,\ \ $\brho\gets\log|\mathbf{y}|-\log|\bbeta|$
  \State $R\gets\clip_{[0,R_{\max}]}\big(-\mathbf{t}^\top\brho/\mathbf{t}^\top\mathbf{t}\big)$
  \Statex \hspace{\algorithmicindent}\textit{(C) Susceptibilities}
  \For{$n=1,\dots,n_{\mathrm{LM}}$}
    \State $\mathbf{s}^\pm\gets c^\pm\exp\!\big((-R+u^\pm\chi^\pm)\mathbf{t}\big)$,\ \ $\mathbf{s}\gets\mathbf{s}^++\mathbf{s}^-+c^0e^{-R\mathbf{t}}$
    \State $\mathbf{r}\gets\big[\log|\mathbf{s}|-\log|\mathbf{y}|,\ \arg(\mathbf{s}\odot\bar{\mathbf{y}})\big]$ \Comment{wrapped residual}
    \State $\mathbf{v}^\pm\gets u^\pm(\mathbf{s}^\pm\oslash\mathbf{s})\odot\mathbf{t}$,\ \ $\mathbf{J}\gets\begin{bmatrix}\Re\mathbf{v}^+&\Re\mathbf{v}^-\\\Im\mathbf{v}^+&\Im\mathbf{v}^-\end{bmatrix}$,\ \ $\bar\sigma\gets\tfrac12\tr(\mathbf{J}^\top\mathbf{J})$
    \State $\bdelta\gets-\big(\mathbf{J}^\top\mathbf{J}+\lambda\bar\sigma\mathbf{I}\big)^{-1}\mathbf{J}^\top\mathbf{r}$ \Comment{$2\times2$, Cramer's rule}
    \State $\bchi_{\mathrm{trial}}\gets\big(\clip_{[0,\chi_{\max}]}(\chi^++\delta^+),\ \clip_{[-\chi_{\max},0]}(\chi^-+\delta^-)\big)^\top$
    \If{$\norm{\mathbf{r}(\bchi_{\mathrm{trial}})}^2<\norm{\mathbf{r}(\bchi)}^2$}
      \State $\bchi\gets\bchi_{\mathrm{trial}}$,\ \ $\lambda\gets\max\{\lambda/\nu,\lambda_{\min}\}$
    \Else
      \State $\lambda\gets\min\{\nu\lambda,\lambda_{\max}\}$
    \EndIf
  \EndFor
  \Statex \hspace{\algorithmicindent}\textit{Selection}
  \State Among the $K$ restarts, track the state minimizing $\mathcal{S}=\norm{\mathbf{s}-\mathbf{y}}^2/\norm{\mathbf{y}}^2+\norm{\mathbf{r}}^2/N_e$
\EndFor
\State With the final stored state and $\Psi(\mathbf{m})=\1^\top\mathcal{U}(\arg\mathbf{m})/(p\,\1^\top\mathbf{t})$:
\Statex \hspace{\algorithmicindent}$\mathrm{PCS}\gets\Psi\big(c^+\mathbf{b}^++(c^0+c^-)\mathbf{b}^0\big)$,\ \ $\mathrm{DCS}\gets\Psi\big(c^-\mathbf{b}^-+(c^0+c^+)\mathbf{b}^0\big)$
\Statex \hspace{\algorithmicindent}$\chi_{\mathrm{comp}}\gets\Psi(\mathbf{B}\mathbf{c})$
\end{algorithmic}
\end{algorithm}

Besides quantities $T$ and $K$, Algorithm~\ref{alg:canon} uses a small number of fixed constants, listed in Table~\ref{tab:params}. With the exception of $\chi_{\max}$, which is chosen per dataset (Sec.~\ref{sec:subproblemC}), they were set once and kept identical in every experiment of this work, at both field strengths and for both the \textit{in silico} and \textit{in vivo} data:

\begin{table}[h]
\caption{Fixed constants of CANON-QSM (Algorithm~\ref{alg:canon}).}\label{tab:params}
\centering\small
\begin{tabular}{|@{}lll@{}|}
\toprule
Symbol & Role & Value\\
\midrule
$T$ & outer alternating iterations & $20$\\
$K$ & restarts per voxel & $20$\\
$n_{\mathrm{LM}}$ & LM steps per outer iteration & $10$\\
$\lambda_0$ & initial LM damping & $10^{-2}$\\
$\nu$ & damping growth/shrink factor & $2$\\
$\lambda_{\min},\,\lambda_{\max}$ & damping limits & $10^{-7},\,10^{7}$\\
$c_f^{(1)}\to c_f^{(T)}$ & concentration floor& $2\times10^{-2}\to10^{-4}$\\
$R_{\max}$ & upper bound of the baseline relaxation rate & $2000$\,s$^{-1}$\\
$\chi_{\max}$ & susceptibility bound & per dataset ($0.48$\,ppm \textit{in silico})\\
\bottomrule
\end{tabular}
\end{table}

\end{document}